\documentclass[conference]{IEEEtran}
\IEEEoverridecommandlockouts
\makeatletter
\def\endthebibliography{%
  \def\@noitemerr{\@latex@warning{Empty `thebibliography' environment}}%
  \endlist
}
\makeatother
\usepackage{cite}
\usepackage{amsmath,amssymb,amsfonts}
\usepackage{dsfont}
\usepackage{algorithm}
\usepackage{algpseudocode}
\usepackage{graphicx}
\usepackage{textcomp}
\usepackage{xcolor}
\usepackage{float}
\usepackage{subfig}
\usepackage{enumitem}

\ifodd 1
\newcommand{\com}[1]{{\color{blue}#1}} 
\else

\newcommand{\com}[1]{}
\fi

\def\BibTeX{{\rm B\kern-.05em{\sc i\kern-.025em b}\kern-.08em
    T\kern-.1667em\lower.7ex\hbox{E}\kern-.125emX}}

\begin{document}

\font\myfont=cmr12 at 21.5pt
\title{{\myfont From Raw Data to Shared 3D Semantics: Task-Oriented Communication for Multi-Robot Collaboration}\\
}
\author{
\IEEEauthorblockN{Ruibo Xue,~Jiedan Tan,~Fang Liu,~Jingwen Tong,~Taotao~Wang,~and~Shuoyao Wang}
\IEEEauthorblockA{College of Electronics and Information Engineering, Shenzhen University, Shenzhen, 518060, China\\
Emails: xueruibo624@163.com; 2510044003@mails.szu.edu.cn; \{liuf, eejwentong, ttwang, sywang\}@szu.edu.cn}}

\maketitle

\begin{abstract}
Multi-robot systems (MRS) rely on exchanging raw sensory data to cooperate in complex three-dimensional (3D) environments. However, this strategy often leads to severe communication congestion and high transmission latency, significantly degrading collaboration efficiency. This paper proposes a decentralized task-oriented semantic communication framework for multi-robot collaboration in unknown 3D environments. Each robot locally extracts compact, task-relevant semantics using a lightweight Pixel Difference Network (PiDiNet) with geometric processing. It shares only these semantic updates to build a task-sufficient 3D scene representation that supports cooperative perception, navigation, and object transport. Our numerical results show that the proposed method exhibits a dramatic reduction in communication overhead from $858.6$ Mb to $4.0$ Mb (over $200\times$ compression gain) while improving collaboration efficiency by shortening task completion from $1,054$ to $281$ steps.
\end{abstract}


\section{Introduction}
Multi-robot systems (MRS) are increasingly deployed in safety-critical applications closely tied to daily life, including autonomous warehouse management, urban search and rescue, and large-scale environmental monitoring~\cite{li2025task}. In these scenarios, multiple robots operate in complex \emph{unknown} three-dimensional (3D) environments and must share situational awareness to coordinate perception, navigation, and execution.

However, extending multi-robot collaboration to unknown 3D environments introduces unique challenges that fundamentally shift the communication bottleneck. 
First, the lack of prior environmental knowledge imposes a dual burden: unlike scenarios with pre-built maps where coordination relies solely on sharing low-bandwidth robot states (e.g., poses and goals), robots in unknown environments must continuously sense, exchange, and collaboratively map the surroundings to avoid redundant exploration and ensure safety~\cite{hughes2022hydra}. 
Second, the transition to 3D environments introduces a ``\emph{curse of dimensionality}.'' Unlike 2D planar mapping, 3D perception generates massive volumetric data (e.g., dense point clouds or TSDF grids) to represent spatial occupancy~\cite{tian2022kimera}. 

While transmitting this raw sensory data preserves rich background information, it is highly inefficient for conveying the geometric constraints actually needed for navigation. When the number of robots increases, the aggregate traffic from these high-fidelity streams saturates the channel, inducing congestion and queuing delays~\cite{luo2022semoverview}. Such coordination degrades not only because fewer bits are delivered, but because the structural updates required for collision avoidance arrive too late for decentralized decision-making.

This limitation reflects a fundamental \emph{data--decision mismatch}. In many collaborative tasks, robots do not need the full texture and density of raw measurements; rather, they need compact, decision-critical facts for spatial alignment~\cite{shi2021semaware}. Motivated by this, \emph{task-oriented communication} has emerged as an alternative to data-centric transmission by prioritizing task-relevant information~\cite{shao2021learning,gunduz2023beyondbits}. Instead of periodically broadcasting high-dimensional data, it delivers compact representations that directly support downstream control~\cite{xie2022task}. For example, in cooperative exploration, continuously streaming raw observations wastes bandwidth on static details, whereas time-critical events (e.g., discovering a new obstacle boundary) should be prioritized. Therefore, event-driven transmission can align communication with these \emph{structural discovery points}, reducing unnecessary traffic~\cite{nowzari2019event}.

However, existing works mainly focus on semantic awareness at the network level~\cite{gong2025digital, shi2021semaware}, limiting their direct applicability to MRS under complex geometric constraints. Recent methods typically emphasize representation efficiency for recognition tasks~\cite{shao2021learning}, without explicitly addressing how exchanged semantics should be organized to support \emph{decentralized} 3D geometric coordination~\cite{xie2022task,xu2023tosa_uav}. Furthermore, many MRS solutions rely on centralized perception pipelines or consistent dense map fusion~\cite{tian2022kimera}, which are computationally heavy in bandwidth-constrained unknown environments.

In this article, we propose a \emph{geometry-aware decentralized task-oriented semantic communication framework} specifically tailored for multi-robot collaboration in unknown 3D environments. Recognizing that \textbf{geometric boundaries (edges)} carry the most critical information for 3D navigation, we design a structural semantic extraction module based on a lightweight Pixel Difference Network (PiDiNet)~\cite{su2021pixel}. This module extracts sparse \textbf{geometric primitives} from raw sensory data, discarding redundant volumetric information. By exchanging these updates via an \textbf{event-triggered mechanism}, which broadcasts only when significant structural changes are detected, robots collaboratively construct a lightweight shared 3D semantic scene. This incremental scene construction allows the team to discover and synchronize the environment layout from scratch without requiring centralized perception. Numerical results demonstrate that the proposed framework substantially reduces communication overhead from $858.6$~Mb to $4.0$~Mb (exceeding $200\times$ compression gain) while accelerating task completion from $1,054$ to $281$ steps.

The main contributions of this paper are listed below:
\begin{itemize}
    \item We develop a decentralized communication paradigm that replaces raw volumetric streams with compact, task-relevant structural updates, specifically addressing the data surge inherent to unknown environment exploration.

    \item We introduce a lightweight 3D scene representation method that fuses 2D semantic edges into a unified 3D geometric space, allowing robots to maintain spatial consistency with minimal data exchange.

    \item We present an integrated system that couples semantic perception, event-driven communication, and decentralized task execution, enabling coordination without a centralized controller or global dense map fusion.
\end{itemize}

\begin{figure*}[!h]
  \centering
  \includegraphics[width=0.95\textwidth]{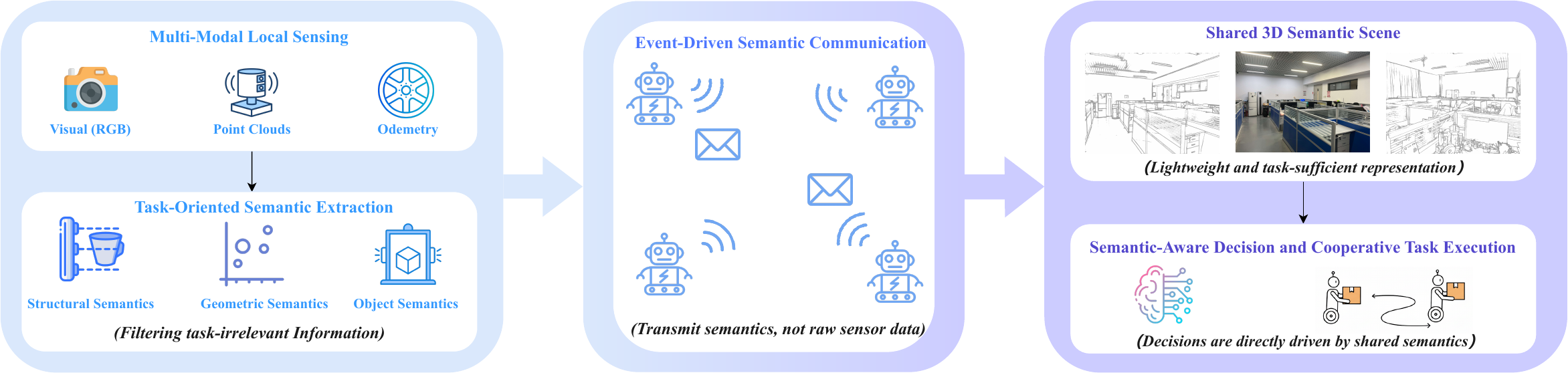}
  \caption{\small Overview of the task-oriented semantic communication framework for multi-robot collaboration. First, multi-modal local sensing data (RGB, point clouds, and odometry) is processed to extract compact structural, geometric, and object semantics, effectively filtering out task-irrelevant information. Second, an event-driven semantic communication mechanism transmits these lightweight semantic packets. Finally, the received data is aggregated into a shared 3D semantic scene, visualized as sparse edge maps, which serves as a lightweight representation to directly drive semantic-aware decision-making and cooperative task execution.}
  \label{fig:framework}
\end{figure*}

\section{System Model and Problem Setup}
\label{sec:system_model}

We consider a team of \(N\) mobile robots operating in an unknown 3D environment.
The global boundary and the unloading depot location are known \emph{a priori}, while internal obstacles and target objects are unknown.
The collective mission is \emph{cooperative search-and-transport}: robots explore, discover targets, pick them up, and deliver them to the depot.

\subsection{Robot State, Observation, and Local Navigation}
\label{subsec:state_obs}

Time is discretized into steps indexed by \(t\).
Robot \(i \in \{1,\dots, N\}\) maintains a local state estimate \(x_t^{(i)}\) (e.g., pose and velocity) from onboard odometry.
At each step, it receives a multimodal observation \(o_t^{(i)}\) (e.g., RGB-D images and/or point clouds).
Each robot performs collision checking and short-horizon motion control locally. Note that the proposed framework does not require sharing raw observations or building a globally consistent dense map.

\subsection{Wireless Communication Model}
\label{subsec:comm_model}

Robots communicate over a bandwidth-limited shared wireless channel~\cite{tong2022age}.
Let \(B_{\max}\) denote the maximum \emph{aggregate} number of transmittable bits per step.
Denote by \(M_t^{(i)}\) the message transmitted by robot \(i\) at step \(t\), with length \(\lvert M_t^{(i)} \rvert\) in bits.
The network capacity constraint is given by
\begin{equation}
    \sum_{i=1}^{N} \lvert M_t^{(i)} \rvert \le B_{\max}, \quad \forall t.
    \label{eq:capacity}
\end{equation}

For comparison, the sensing-centric baseline periodically transmits raw sensory payloads of size \(S_{\mathrm{raw}}\), which scales with sensor dimensionality.
In contrast, our framework transmits compact semantic updates of size \(S_{\mathrm{sem}}(t)\), whose magnitude is decoupled from raw observation size and primarily depends on task-relevant events.

\subsection{Problem Statement}\label{subsec:problem_statement}

The system's goal is to complete the cooperative search-and-transport mission as efficiently as possible under the bandwidth constraint in~\eqref{eq:capacity}. Let $\mathcal{T}$ denote the task utilities.
One needs to find a decentralized policy that (i) extracts task-sufficient semantics from local observations, (ii) schedules semantic transmissions under bandwidth limits, and (iii) enables distributed coordination using only exchanged semantics.
Mathematically, the objective can be expressed as
\begin{equation}
    \min \; \mathcal{T} \quad \text{s.t.} \quad \eqref{eq:capacity},
    \label{eq:objective}
\end{equation}
where the constraint captures the key communication bottleneck in multi-robot collaboration. To solve this problem, we propose the task-oriented semantic communication, as shown in Fig.~\ref{fig:framework}, which consists of three main processing blocks.

\section{Task-Oriented Semantic Perception and Event-Driven Communication}
\label{sec:semantic_comm}

This section introduces the left and middle blocks of Fig.~\ref{fig:framework}.
Each robot extracts compact 3D task-oriented semantics from local multimodal observations and transmits \emph{event-driven} semantic updates under strict bandwidth constraints.
No centralized fusion or decision logic is assumed in this stage.

\subsection{Task-Oriented 3D Semantic Perception}
\label{sec:semantic_perception}

From the local observation \(o_t^{(i)}\), robot \(i\) constructs three complementary semantic components:

\textbf{1) Structural semantics (edges).}
We run PiDiNet on the onboard RGB image to obtain an edge probability map, which is then thresholded and downsampled to a binary edge mask \(E_t^{(i)}\)  that captures the scene layout and obstacle boundaries.
This representation preserves topological constraints while discarding appearance details irrelevant to the task.

\textbf{2) Geometric semantics (sparse anchors).}
We extract sparse 3D anchors \(P_{\mathrm{sparse},t}^{(i)}\) from depth/point clouds by voxel downsampling and outlier filtering.
These anchors preserve coarse spatial constraints and support cross-robot spatial alignment at low bandwidth.

\textbf{3) Object semantics (target entities).}
Detected targets are abstracted into semantic entities
\begin{equation}
    s_k = \big(\mathrm{id}_k,\, l_k,\, p_k,\, \theta_k,\, \sigma_k,\, \mathrm{st}_k\big),
    \label{eq:entity}
\end{equation}
where \(\mathrm{id}_k\) is a unique entity identifier, \(l_k\) is the category label, \(p_k \in \mathbb{R}^3\) is the estimated 3D position, \(\theta_k\) is the orientation, \(\sigma_k\) is a confidence score, and \(\mathrm{st}_k\) is a discrete task status (e.g., unassigned/claimed/carried/delivered).

The above three components jointly capture \emph{(i)} traversability and obstacle boundaries (edges), \emph{(ii)} coarse geometry for alignment (anchors), and \emph{(iii)} task-relevant goals and states (objects), which is sufficient for decentralized coordination.

\subsection{Semantic Message Format and Bit Budget}
\label{sec:semantic_format}

At time step $t$, robot $i$ encodes its semantics into a compact message
\begin{equation}
    M_{\mathrm{sem},t}^{(i)} = \Big\{ H_t^{(i)},\, E_t^{(i)},\, P_{\mathrm{sparse},t}^{(i)},\, O_t^{(i)} \Big\},
    \label{eq:semantic_msg}
\end{equation}
where \(H_t^{(i)}\) is a lightweight header (robot ID, timestamp, and optional pose), \(E_t^{(i)}\) is the binary edge mask, \(P_{\mathrm{sparse},t}^{(i)}\) is the sparse anchor set, and \(O_t^{(i)}=\{s_k\}\) is the set of object entities.
To ensure the communication cost is explicit and reproducible, we adopt the following practical encoding choices:
\begin{itemize}
    \item \textbf{Edge payload:} a \(256\times256\) 1-bit mask, i.e., \(\lvert E_t^{(i)} \rvert \approx 65,536\) bits (\(\approx 64\,\mathrm{Kb}\)).
    \item \textbf{Anchor payload:} \(m\) anchors with fixed-point quantized coordinates and confidence.
    Using 48 bits per anchor and \(m \approx 1200\), we obtain \(\lvert P_{\mathrm{sparse},t}^{(i)} \rvert \approx 5.8\times10^4\) bits.
    \item \textbf{Object payload:} up to \(K\) entities with compact fields (ID, label, quantized pose, confidence, status).
    For \(K \le 20\), the object payload is typically a few kilobits.
\end{itemize}

In summary, a semantic update has a nominal payload on the order of \({O}(10^2)\,\mathrm{Kb} \), remaining well below raw sensing payloads while preserving task-critical information.

\subsection{Event-Driven Transmission Policy}
\label{sec:event_driven}
As given in Algorithm~\ref{alg:event}, robots broadcast semantic updates only when \emph{task-relevant} changes occur to avoid periodic high-rate transmissions.
Let \(\gamma_t^{(i)} \in \{0,1\}\) denote a transmission indicator for robot \(i\) at step \(t\).
We transmit if any of the following events are triggered:

\noindent \textbf{(E1) New target:} A transmission is triggered if a new object $s_k$ is detected in the current observation set ${O}_t^{(i)}$:
\begin{equation}
    \exists s_k \in {O}_t^{(i)} \quad \text{s.t. } s_k \text{ is newly discovered}.
    \label{eq:event_new_target}
\end{equation}

\noindent \textbf{(E2) Pose refinement:} An update is broadcast if the geometric estimate of object $k$ deviates significantly from its last transmitted state:
\begin{equation}
    \lVert p_k(t) - \hat{p}_k(t^-)\rVert > \delta_p \quad \lor \quad |\theta_k(t) - \hat{\theta}_k(t^-)| > \delta_\theta,
    \label{eq:event_pose}
\end{equation}
where $\delta_p$ and $\delta_\theta$ are the thresholds for the deviations of the robot position and orientation, respectively.

\noindent \textbf{(E3) Status change:} A message is sent immediately if the discrete task status of object $k$ changes:
\begin{equation}
    \mathrm{st}_k(t) \neq \mathrm{st}_k(t^-).
    \label{eq:event_status}
\end{equation}

\noindent \textbf{(E4) Structural change:} 3D scene updates are triggered when the local edge map changes significantly:
\begin{equation}
    d\big(E_t^{(i)}, E_{t^-}^{(i)}\big) > \delta_E.
    \label{eq:event_structure}
\end{equation}

Hence, we have $\gamma_t^{(i)} = 1$, if any event in \textbf{E1}--\textbf{E4} holds.

Finally, we employ a lightweight priority rule to satisfy the shared-channel limit in~\eqref{eq:capacity}:
object-status changes (\textbf{E3}) and new targets (\textbf{E1}) are highest priority, followed by pose refinements (\textbf{E2}), and then structural updates (\textbf{E4}) if budget remains.

\begin{algorithm}[t]
\caption{Event-driven semantic generation at robot \(i\)}
\label{alg:event}
\begin{algorithmic}[1]
\State Observe \(o_t^{(i)}\), update pose estimate \(x_t^{(i)}\)
\State Run PiDiNet to obtain \(E_t^{(i)}\); extract anchors \(P_{\mathrm{sparse},t}^{(i)}\); detect objects \(O_t^{(i)}\)
\State Evaluate events \textbf{E1}--\textbf{E4} and set \(\gamma_t^{(i)}\) by \eqref{eq:event_new_target}--\eqref{eq:event_structure}
\If{\(\gamma_t^{(i)}=1\)}
    \State Encode \(M_{\mathrm{sem},t}^{(i)}\) by \eqref{eq:semantic_msg} and broadcast subject to \eqref{eq:capacity}
\EndIf
\end{algorithmic}
\end{algorithm}

\section{Shared 3D Semantic Scene and Decentralized Cooperation}
\label{sec:decision}
This section introduces the right block of Fig.~\ref{fig:framework}.
By using only exchanged semantic messages, each robot incrementally builds a lightweight shared semantic scene that is sufficient for decentralized task allocation, motion planning guidance, and cooperative transport execution.

\subsection{Shared 3D Semantic Scene Representation}
\label{sec:shared_scene}
By aggrating the compact message $M^{j}_{\mathrm{sem},t}$ from different robots,
each robot \(i\) maintains a local shared semantic scene
\begin{equation}
    \mathcal{S}_t^{(i)} = \big(\mathcal{E}_t^{(i)},\, \mathcal{P}_t^{(i)},\, \mathcal{O}_t^{(i)}\big),
    \label{eq:shared_scene}
\end{equation}
where \(\mathcal{E}_t^{(i)}\) stores aggregated structural edges, \(\mathcal{P}_t^{(i)}\) stores merged sparse anchors, and \(\mathcal{O}_t^{(i)}\) stores the synchronized object/entity table (including statuses).
Upon receiving a message \(M_{\mathrm{sem},t}^{(j)}\) from robot \(j\), robot \(i\) fuses it into \(\mathcal{S}_t^{(i)}\) based on:

\textbf{1) Anchor merging.}
Anchors are merged by voxel hashing in 3D: anchors falling into the same voxel are averaged (or the highest-confidence anchor is kept), yielding a bounded memory footprint and robustness to duplicate reports.

\textbf{2) Object data association and synchronization.}
For an incoming entity \(s_k\), we match it to an existing entity \(s_{k'}\in\mathcal{O}_t^{(i)}\) if labels agree and the distance is within a gate:
\begin{equation}
    \lVert p_k - p_{k'} \rVert \le \epsilon_o.
    \label{eq:assoc}
\end{equation}
Matched entities are fused by confidence-weighted averaging, while statuses \(\mathrm{st}_k\) are updated with a monotone rule (e.g., \(\text{delivered} \succ \text{carried} \succ \text{claimed} \succ \text{unassigned}\)) to avoid oscillation under asynchronous updates.

\textbf{3) Structural edge aggregation.}
Edge masks are used as coarse traversability constraints.
We maintain a conservative union (or confidence-weighted sum with thresholding) over received edge masks after applying the sender's reported pose transform (when available), producing \(\mathcal{E}_t^{(i)}\) that captures obstacle boundaries at a task-sufficient granularity.

Importantly, \(\mathcal{S}_t^{(i)}\) is \emph{not} intended to be a globally consistent dense reconstruction.
Instead, it is a compact, incrementally synchronized 3D semantics for multi-robot collaboration.

\subsection{Decentralized Semantic-Aware Task Allocation}
\label{sec:task_allocation}

Task allocation is performed in a fully decentralized manner using the shared object table \(\mathcal{O}_t^{(i)}\).
When an entity is in the \emph{unassigned} state, each robot computes a local cost
\begin{equation}
    c_{ik} = \alpha \, \lVert p_k - p^{(i)}_t \rVert + \beta \, \mathbb{I}[\text{robot \(i\) is busy}],
    \label{eq:cost}
\end{equation}
where \(p^{(i)}_t\) is robot \(i\)'s current position, and \(\alpha,\beta\) are weights.
The robot with the lowest cost claims the target by broadcasting a lightweight \emph{lock/claim} message (containing entity ID, claimer ID, and timestamp).
Upon receiving it, other robots mark the target as \emph{claimed} and refrain from redundant pursuit.

\subsection{Semantic Motion Planning and Collision Avoidance}
\label{sec:planning}

Each robot plans and navigates primarily using local sensing, while the shared semantic scene provides global guidance. First, \(\mathcal{E}_t^{(i)}\) and \(\mathcal{P}_t^{(i)}\) define coarse non-traversable regions and boundary hints, improving exploration safety and reducing repeated dead-ends. Second, object entities in \(\mathcal{O}_t^{(i)}\) provide goal locations for assigned targets and enable team-wide awareness of remaining tasks. Third, teammate states implied by claims/statuses are used to reduce interference; robots treat teammates as dynamic obstacles when planning local trajectories. Finally, by planning with task-level semantics rather than dense globally consistent maps, the system remains robust to partial/asynchronous updates and incurs low computation and communication overhead.

\subsection{Cooperative Transport Execution}
\label{sec:execution}

Once robot \(i\) reaches its assigned target, it performs pickup and updates the corresponding entity status \(\mathrm{st}_k\) to \emph{carried} by broadcasting a short status message.
The object is then transported to the depot.
After successful delivery, the target is marked as \emph{delivered} in \(\mathcal{O}_t^{(i)}\), and this status is disseminated to the team via the event-driven mechanism in Section~\ref{sec:event_driven}.
This completes the perception--communication--decision--action loop under strict bandwidth constraints.

\begin{figure*}[t]
  \centering
  \includegraphics[width=0.94\textwidth]{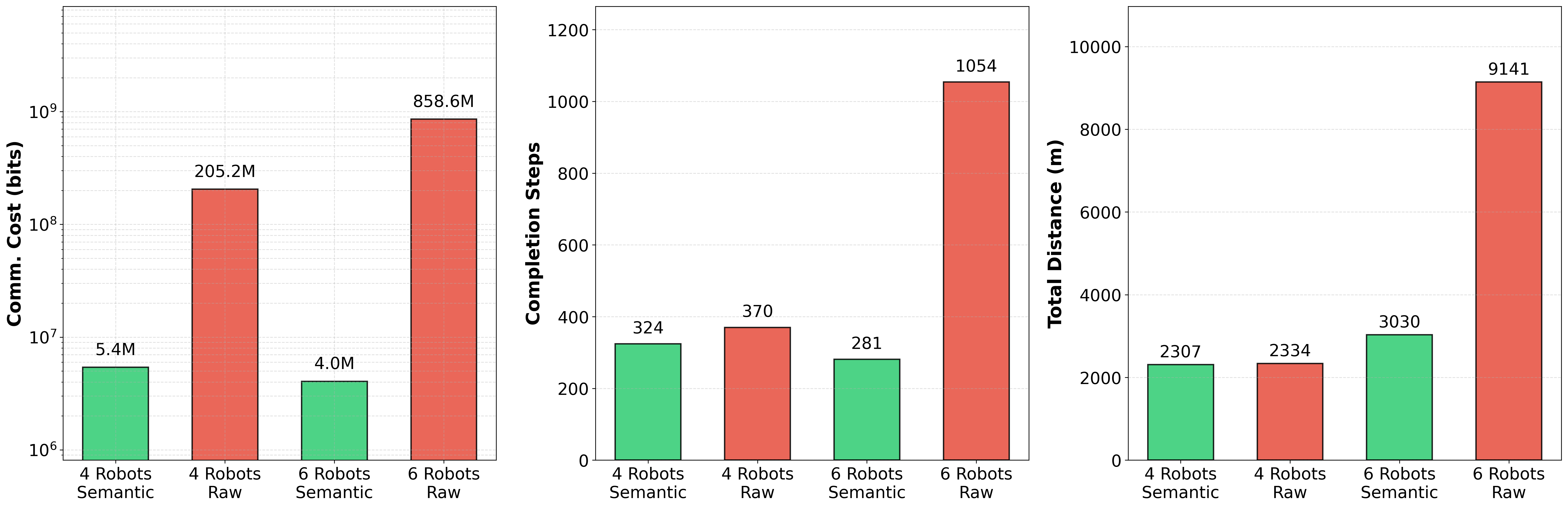}
 \caption{\small The performance comparison between raw and semantic communication under bandwidth constraints: (a) communication cost, (b) task completion steps, and (c) total traveled distance.}
  \label{fig:system_performance}
\end{figure*}

\begin{figure}[t]
  \centering
  \includegraphics[width=0.5\textwidth]{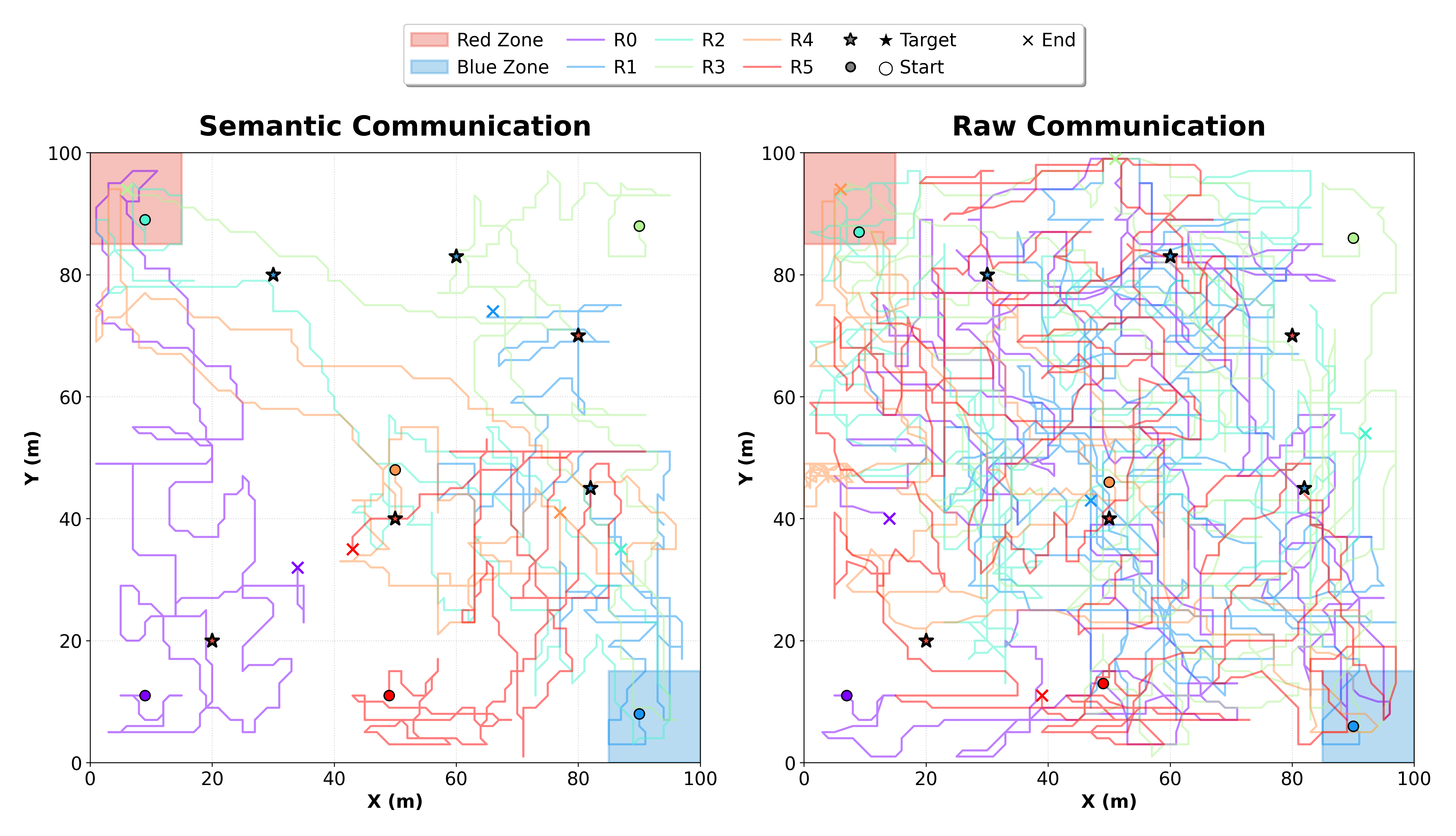}
  \caption{\small A visualization of robot trajectories under semantic communication (left) and raw communication (right).}
  \label{fig:trajectories}
\end{figure}

\begin{figure}[t]
  \centering
  \includegraphics[width=1\linewidth]{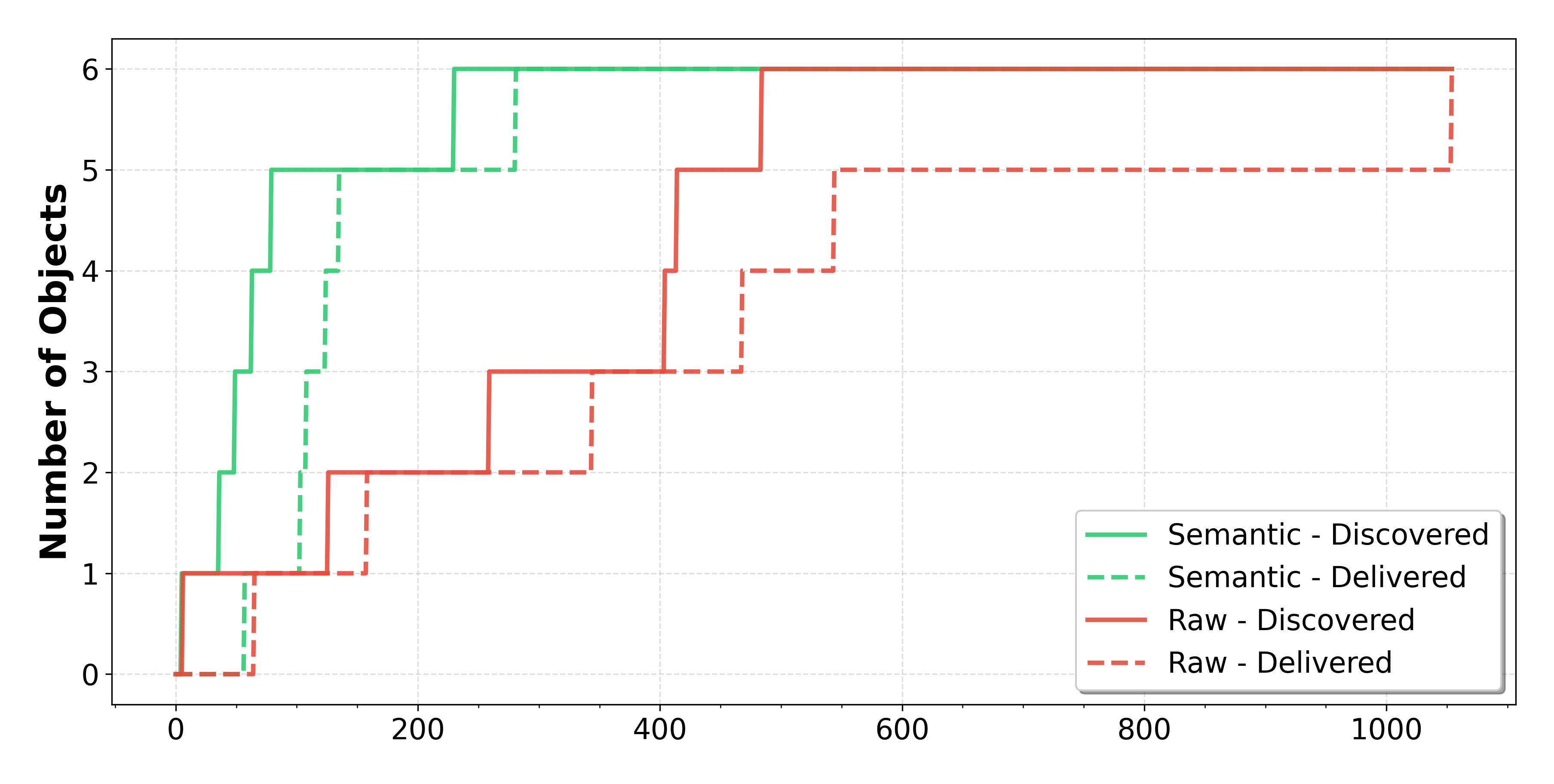}
  \caption{\small The cumulative number of discovered and
delivered targets.}
  \label{fig:timeline}
\end{figure}

\begin{figure*}[t]
  \centering
  \includegraphics[width=0.9\textwidth]{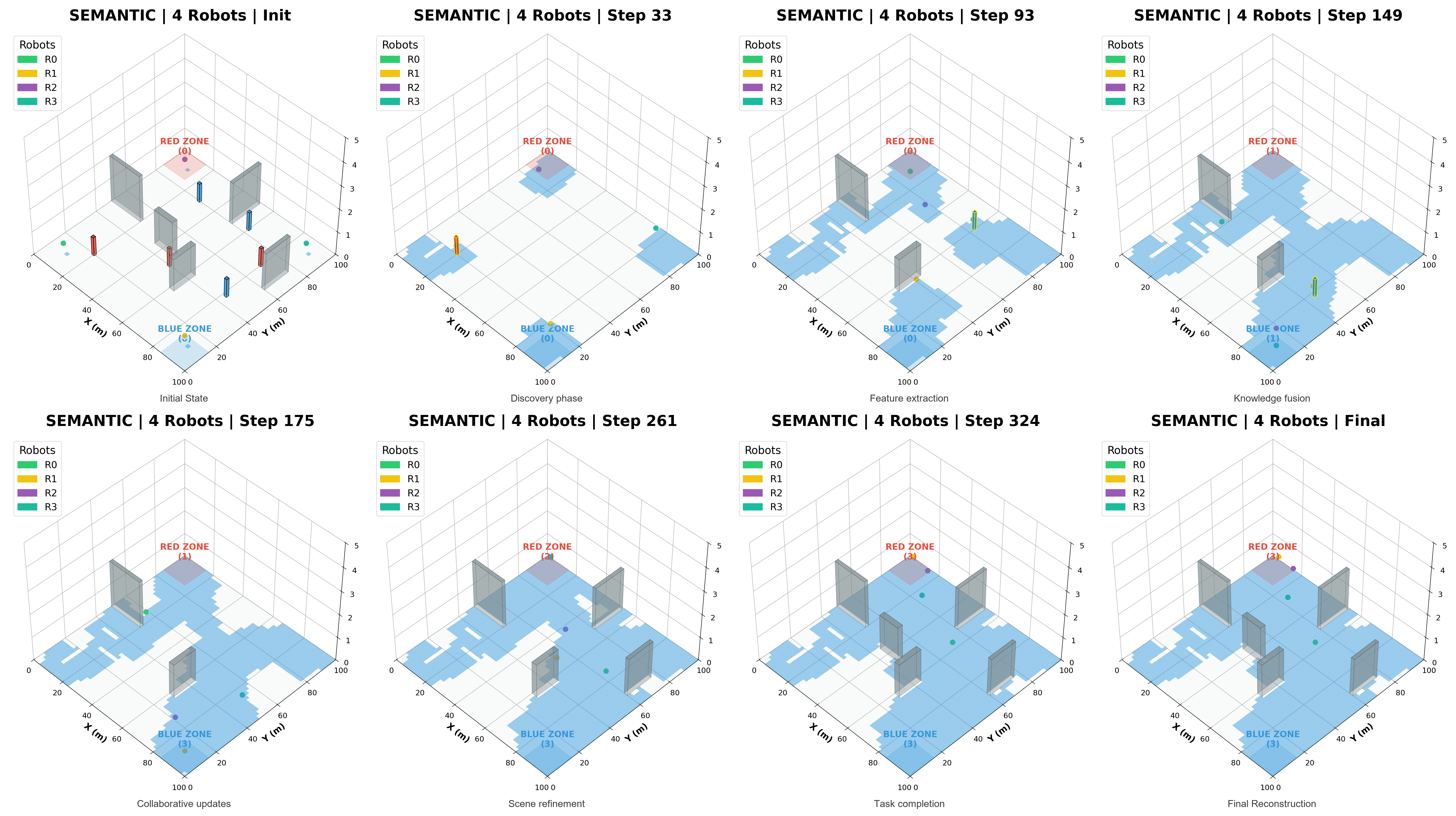}
\caption{\small An illustration of the shared 3D semantic scene evolving in the four-robot semantic setting, from initialization to the final state. It incrementally constructs the shared task-oriented semantic scene enabled by event-driven semantic communication.}
\label{fig:reconstruction}
\end{figure*}

\section{Experiments}
\label{sec:experiments}

\subsection{Experimental Setup}
\label{subsec:exp_setup}
We evaluate the proposed framework in a simulated multi-robot search-and-transport task under strict bandwidth constraints. Unless otherwise stated, the team consists of four robots, and we additionally test a six-robot configuration to examine scalability. Robots operate in an initially unknown 3D environment with static obstacles and multiple target objects. The mission is to explore, discover targets, and deliver them to a designated depot. A grid map is used \emph{only} inside the simulator for collision checking and task planning, while inter-robot coordination relies on communicated information.

We compare two communication paradigms over a shared wireless channel: (i) \textbf{raw communication}, which periodically transmits high-dimensional sensing-centric observations, and (ii) \textbf{semantic communication (ours)}, which transmits compact, task-relevant semantic updates in an event-driven manner. All other modules (decentralized task allocation policy, exploration strategy, and motion planning) are kept the same to isolate the effect of communication.

We report three system-level metrics: \textbf{(a)} total communication cost (transmitted bits), \textbf{(b)} task completion steps, and \textbf{(c)} total traveled distance. In addition, we analyze \textbf{discovery-to-delivery latency} to quantify coordination responsiveness, and provide qualitative visualizations of trajectories and shared semantic scene construction.

\subsection{Results and Analysis}
\label{subsec:results}

Fig.~\ref{fig:system_performance} summarizes the system-level comparison between raw and semantic communication.

\textbf{Communication cost (Fig.~\ref{fig:system_performance}(a)).}
Semantic communication drastically reduces bandwidth usage in both team sizes. In the four-robot setting, the total communication cost decreases from {205.2~Mb} (raw) to {5.4~Mb} (semantic), i.e., about {38$\times$} reduction. In the six-robot setting, the gap becomes even larger: raw communication increases to \textbf{858.6~Mb}, while semantic communication remains at only \textbf{4.0~Mb}, yielding over \textbf{200$\times$} reduction. 
This reveals that, by transmitting \emph{task-level semantics} rather than high-dimensional sensor streams, communication load is largely decoupled from sensing modality size and scales much more gracefully with team size.

\textbf{Task completion steps (Fig.~\ref{fig:system_performance}(b)).}
Reduced congestion translates into faster global progress. With four robots, semantic communication completes the task in {324} steps versus {370} steps for raw communication. The advantage becomes pronounced with six robots: semantic communication finishes in \textbf{281} steps, while raw communication requires \textbf{1,054} steps (approximately \textbf{3.8$\times$} slower). 
This indicates that under tight bandwidth, raw communication suffers from severe information latency and coordination delays as the team grows, whereas event-driven semantic updates preserve timely synchronization of task states and intentions.

\textbf{Total traveled distance (Fig.~\ref{fig:system_performance}(c)).}
Semantic communication consistently reduces redundant motion. In the four-robot case, the total traveled distance is similar ({2,307~m} semantic vs. {2,334~m} raw), suggesting both methods can complete the task with comparable motion efficiency at a small scale. However, in the six-robot case, semantic communication significantly reduces distance from {9,141~m} (raw) to {3,030~m} (semantic), i.e., about {3$\times$} reduction. 

Fig.~\ref{fig:trajectories} provides qualitative trajectory visualizations. Under semantic communication (left), robot paths are more spatially separated with less overlap, indicating a clearer implicit division of exploration regions and fewer repeated visits. Under raw communication (right), trajectories are densely intertwined across the environment, with frequent crossings and revisits, which is consistent with the substantially larger traveled distance observed in the six-robot raw setting (Fig.~\ref{fig:system_performance}(c)). This visual evidence supports the conclusion that task-oriented semantic synchronization reduces redundant exploration and improves team-level efficiency.

Fig.~\ref{fig:timeline} plots the cumulative number of discovered and delivered targets over time. Semantic communication exhibits a noticeably smaller gap between the discovery curve (solid) and delivery curve (dashed), meaning targets are converted into completed deliveries more quickly after being found. In contrast, raw communication shows a larger and longer-lasting separation between discovery and delivery, reflecting delayed task claiming/assignment and slower coordinated transport.
This result directly echoes our mechanism-level argument: compact, event-triggered messages keep the shared task state fresh, enabling faster transitions from perception events to coordinated actions under bandwidth limits.

Fig.~\ref{fig:reconstruction} illustrates how the \textbf{Shared 3D Semantic Scene} evolves in the four-robot semantic setting, from initialization to the final state. As exploration proceeds, robots progressively contribute local discoveries and task-relevant structural cues; the shared scene becomes increasingly complete in a step-by-step manner, enabling consistent task awareness and coordination. Importantly, this shared representation is designed to be \emph{task-sufficient} rather than metrically perfect: the goal is not a globally consistent mapping, but maintaining the semantic state needed to support decentralized decisions. The steady enrichment of the shared scene throughout the task provides a concrete qualitative explanation for the improved responsiveness and scalability observed in Figs.~\ref{fig:system_performance}--\ref{fig:timeline}.

\subsection{Discussion and Analysis}
\label{subsec:discussion}
Across quantitative metrics (communication cost, completion steps, traveled distance) and qualitative evidence (timeline and trajectories), the results consistently validate that: \textbf{aligning communication with task-level semantics yields tightly coupled gains in bandwidth efficiency and coordination quality under resource constraints}. Raw communication becomes increasingly ineffective as team size grows because high-volume transmissions induce congestion and stale information, which amplifies coordination conflicts and redundant motion. In contrast, event-driven semantic communication maintains timely synchronization of \textbf{decision-critical} information, enabling scalable cooperation with lower bandwidth usage and substantially improved task efficiency in larger teams.

\section{Conclusion}\label{sec:conclusion}
We presented a task-oriented semantic communication framework that decouples multi-robot coordination from high-dimensional sensing data. By transmitting only event-driven semantic updates, our approach reduces bandwidth consumption by over $200\times$ and accelerates task completion by nearly $4\times$ in large teams compared to raw communication. Specifically, the maintenance of a task-sufficient shared scene significantly mitigates redundant motion, reducing total traveled distance by approximately $3\times$ while ensuring low latency between target discovery and delivery. These results confirm that aligning information exchange with task utility, rather than data volume, is critical for achieving scalable, responsive collaboration in bandwidth-constrained environments. 
\bibliographystyle{IEEEtran}
\bibliography{references}

\end{document}